\pdfoutput=1

\documentclass[11pt,a4paper]{article}
\usepackage{bm}
\usepackage{jheppub}
\usepackage{mathrsfs}

\let\de=\partial
\let\eps=\epsilon
\DeclareMathOperator{\tr}{Tr}
\newcommand{\imag}{\text{i}}
\newcommand{\La}{\mathscr{L}}
\newcommand{\De}{\mathscr{D}}
\newcommand{\Me}{\mathscr{M}}
\newcommand\dd{\mathop{\text d}\nolimits}
\newcommand{\Bex}{B_\text{ex}}
\newcommand{\he}[1]{{#1}^{\dagger}}
\newcommand{\vek}[1]{\bm{#1}}
\newcommand{\skal}[2]{\vek{#1}\cdot\vek{#2}}
\newcommand{\vekt}[2]{\vek{#1}\times\vek{#2}}
\newcommand{\bra}[1]{\langle{#1}|}
\newcommand{\ket}[1]{|{#1}\rangle}

\title{Anomalous low-temperature thermodynamics of QCD in strong magnetic fields}

\author[a]{Tom\'a\v{s} Brauner}
\author[b]{and Saurabh V.~Kadam}
\affiliation[a]{Faculty of Science and Technology, University of Stavanger,\\
N-4036 Stavanger, Norway}
\affiliation[b]{Indian Institute of Science Education and Research (IISER),\\
Pune 411008, India}
\emailAdd{tomas.brauner@uis.no}
\emailAdd{saurabh.kadam@students.iiserpune.ac.in}

\abstract{The thermodynamics of quantum chromodynamics at low temperatures and in sufficiently strong magnetic fields is governed by neutral pions. We analyze the interacting system of neutral pions and photons at zero baryon chemical potential using effective field theory. As a consequence of the axial anomaly and the external magnetic field, the pions and photons mix with one another. The resulting spectrum contains one usual, relativistic photon state, and two nonrelativistic modes, one of which is gapless and the other gapped. Furthermore, we calculate the leading, one-loop contribution to the pressure of the system. In the chiral limit, a closed analytic expression for the pressure exists, which features an unusual scaling with temperature and magnetic field, $T^3B/f_\pi$, at low temperatures, $T\ll B/f_\pi$. Finally, we determine the pion decay rate as a function of the magnetic field at the tree level. The result is affected by a competition of the anisotropic kinematics and the enlarged phase space due to the anomalous mass of the neutral pion. In the chiral limit, the decay rate scales as $B^3/f_\pi^5$.}

\keywords{Anomalies in Field and String Theories, Phase Diagram of QCD}

\arxivnumber{}

\begin{document}

\maketitle


\section{Introduction}
\label{sec:intro}

Extreme magnetic fields with strengths estimated to reach up to $10^{19}\text{ G}$ can exist in the universe, most notably in the terrestrial relativistic heavy ion collision experiments, and in the interiors of certain neutron stars: the magnetars. Since the energy scale associated with such magnetic fields is comparable to the characteristic scale of strong nuclear interactions, this fact has prompted intensive work on the structure of the phase diagram of quantum chromodynamics (QCD) in presence of magnetic fields (see ref.~\cite{Andersen:2014xxa} for a recent review).

While most of the efforts have focused on the effect of the magnetic field on the chiral order parameter of QCD and the chiral phase transition at high temperature and zero baryon chemical potential, it was noticed early on that a \emph{nonzero} chemical potential combined with the magnetic field leads to a qualitative change of the QCD ground state~\cite{Son:2007ny}. Namely, by virtue of the chiral anomaly, sufficiently strong magnetic fields lead to a spatially nonuniform condensate of neutral pions, which takes the form of a soliton lattice and can be energetically favored over normal nuclear matter~\cite{Yamamoto:2015maz,Brauner:2016pko}. A number of recent publications is devoted to a detailed investigation of this new state of matter~\cite{Ozaki:2016vwu,Qiu:2016hzd,Brauner:2017mui}.

In this paper, we draw inspiration from the above studies and investigate the effects of the chiral anomaly on the low-energy dynamics and low-temperature thermodynamics of QCD in strong magnetic fields and \emph{zero} baryon chemical potential. In this case, the structure of the uniform, chiral-symmetry-breaking QCD vacuum remains unaltered, at least for magnetic fields $B\lesssim m_\rho^2\approx0.6\text{ GeV}^2$~\cite{Chernodub:2011mc}. At the same time, charged pions become heavy as a consequence of the Landau level quantization. Hence, the low-energy physics of QCD in presence of a magnetic field is dominated by neutral pions and photons. This is an ideal system to test the effects of the chiral anomaly: the neutral pions do not couple minimally to the electromagnetic field, and the anomaly thus provides the \emph{only} interaction between the two subsystems.

It is well known that in the vacuum, a neutral pion decays anomalously into two photons. At the effective-field-theory (EFT) level, this process can be described by an interaction term proportional to $\phi\eps^{\mu\nu\alpha\beta}F_{\mu\nu}F_{\alpha\beta}$, where $\phi$ represents the neutral pion and $F_{\mu\nu}\equiv\de_\mu A_\nu-\de_\nu A_\mu$ is the usual electromagnetic field strength tensor. Our key observation is that in presence of a background magnetic field, $\vek\Bex$, the same interaction leads to a mixing between pions and photons. This is detailed in section~\ref{sec:spectrum} upon a brief overview of the EFT setup in section~\ref{sec:EFT}. It turns out that one of the two photon polarizations is insensitive to the presence of the background field and retains its relativistic dispersion relation. It is easy to see that this mode carries electric field perpendicular to $\vek\Bex$, since for such field configurations, $\eps^{\mu\nu\alpha\beta}F_{\mu\nu}F_{\alpha\beta}\propto\skal EB$ vanishes. The other photon polarization mixes with the neutral pion, giving rise to two modes with nonrelativistic and anisotropic dispersion relations. One of the two modes is gapless and, interestingly, its dispersion relation becomes quadratic for directions of propagation perpendicular to $\vek\Bex$.

The following sections then present an analysis of some more directly observable consequences of this anomaly-induced mixing between pions and photons. First, in section~\ref{sec:pressure}, we evaluate the pressure of the system at nonzero temperature in the leading, one-loop approximation. Intriguingly, there is a very simple closed analytic expression for the pressure in the chiral limit (zero pion mass) despite the complicated dispersion relation of the pion-photon modes. In section~\ref{sec:decay}, we then analyze the effect of the magnetic field on the above-mentioned anomalous electromagnetic decay of the neutral pion. Finally, in the concluding section~\ref{sec:summary}, we get back to some of the assumptions implicit to our analysis. First we discuss the separation of scales corresponding to the charged and neutral pion sectors, which defines the range of magnetic fields in which our EFT is applicable. Second, we compare the anomalous contributions to the pion spectrum at the tree level to the normal, one-loop corrections, neglected here.

Our main tool is the chiral perturbation theory~\cite{Gasser:1983yg,Gasser:1984gg,Gasser:1987ah}, which governs the dynamics of QCD at low energies and temperatures. Its predictions for observables are organized in a derivative expansion, controlled by the parameter $p/(4\pi f_\pi)$, where $f_\pi\approx92\text{ MeV}$ is the pion decay constant and $p$ the characteristic momentum scale of the system. This limits the validity of our results to temperature and magnetic field scales well below the scale of chiral symmetry breaking. Moreover, we resort to the lowest-order approximations appropriate for the given physical observable. Thus, the excitation spectrum in section~\ref{sec:spectrum} and the pion decay in section~\ref{sec:decay} are analyzed at tree level, and thus include no feedback from the thermal medium. The result for the pressure presented in section~\ref{sec:pressure} then corresponds to a gas of free quasiparticles with dispersion relations fixed to their zero-temperature values.

Throughout the paper, we use the natural units in which the Planck constant $\hbar$, speed of light $c$ as well as the elementary electric charge $e$ are all set to one. The magnetic field strength is given in the high-energy-physics units of $1\text{ GeV}^2\approx1.7\times10^{20}\text{ G}$.


\section{Low-energy effective theory}
\label{sec:EFT}

 The leading order of the chiral perturbation theory Lagrangian for two quark flavors reads
\begin{equation}
\La_\text{$\chi$PT}=\frac{f_\pi^2}4\left[\tr(D_\mu\he\Sigma D^\mu\Sigma)+m_\pi^2\tr(\Sigma+\he\Sigma)\right].
\label{ChPT}
\end{equation}
Here $\Sigma$ is the unimodular and unitary $2\times2$ matrix field that includes the three physical pion degrees of freedom and $m_\pi$ is the pion mass. For the neutral pion, the physical value is $m_\pi\approx135\text{ MeV}$. The minimal coupling of the charged pions to the electromagnetic field $A_\mu$ is introduced through the covariant derivative, $D_\mu\Sigma\equiv\de_\mu\Sigma-\imag[Q_\mu,\Sigma]$, where $Q_\mu\equiv A_\mu\frac{\tau_3}2$, and $\tau_3$ is the third Pauli matrix.

Throughout this paper, we will restrict ourselves to the neutral pion degree of freedom $\phi$, that is, replace $\Sigma\to\exp\bigl(\frac\imag{f_\pi}\phi\tau_3\bigr)$. The master Lagrangian that forms the basis of all the subsequent arguments, is then given by
\begin{equation}
\La=\frac12(\de_\mu\phi)^2+m_\pi^2f_\pi^2\cos\frac{\phi}{f_\pi}-\frac C8\phi\epsilon^{\mu\nu\alpha\beta}F_{\mu\nu}F_{\alpha\beta}-\frac14F_{\mu\nu}F^{\mu\nu}-\frac1{2\xi}(\de_\mu A^\mu)^2.
\label{master}
\end{equation}
The first two terms arise from the chiral perturbation theory Lagrangian~\eqref{ChPT}. The third term stems from the anomalous Wess-Zumino-Witten coupling of $\Sigma$ to a background electromagnetic field and takes the given simple form when restricted to the neutral pion degree of freedom~\cite{Son:2007ny}; we have defined
\begin{equation}
C\equiv\frac1{4\pi^2f_\pi}
\end{equation}
for the sake of brevity. Finally, the last two terms in the effective Lagrangian~\eqref{master} introduce the dynamical electromagnetic field including the standard gauge-fixing term.

In the conventional derivative expansion, the pion field $\phi$ itself counts as order zero, whereas any derivatives acting on it count as order one, and so does the pion mass $m_\pi$. In contrast to the more common way of counting the powers associated with the electromagnetic field, we will treat $A_\mu$ as an object of order zero just like $\phi$. Note that this is consistent with gauge invariance thanks to the fact that we only consider neutral pions here, and also expresses the fact that we are looking at pion dynamics in strong background fields. The master Lagrangian~\eqref{master} then represents the complete effective Lagrangian at the leading, second order of the derivative expansion. This way, including the Wess-Zumino-Witten term in the leading-order Lagrangian is made consistent with power counting.

For the record, we write down explicitly the equations of motion following from the Lagrangian~\eqref{master}. In the covariant relativistic notation, the equations for $\phi$ and $A_\mu$ read
\begin{equation}
\begin{split}
\Box\phi+m_\pi^2f_\pi\sin\frac\phi{f_\pi}+\frac C8\eps^{\mu\nu\alpha\beta}F_{\mu\nu}F_{\alpha\beta}&=0,\\
\Box A^\mu-\left(1-\frac1\xi\right)\de^\mu\de_\nu A^\nu-\frac C2\eps^{\mu\nu\alpha\beta}(\de_\nu\phi)F_{\alpha\beta}&=0.
\end{split}
\label{EoM_rel}
\end{equation}
In order to find the physical interpretation of the various modes in the spectrum, it will also be convenient to have at hand the nonrelativistic, gauge-invariant form of the classical equations of motion. Dropping the gauge-fixing term and trading the field strength tensor $F^{\mu\nu}$ for the electric and magnetic field $\vek E$ and $\vek B$, the equations of motion take the form
\begin{align}
\notag
\ddot\phi-\vek\nabla^2\phi+m_\pi^2f_\pi\sin\frac\phi{f_\pi}&=C\skal EB,\\
\label{EoM_NR}
\skal\nabla E&=-C\skal B\nabla\phi,\\
\notag
\vekt\nabla B&=\dot{\vek E}+C\vek B\dot\phi-C\vekt E\nabla\phi,
\end{align}
where dots stand for time derivative. These are the equations of axion electrodynamics~\cite{Wilczek:1987mv}.


\section{Excitation spectrum}
\label{sec:spectrum}

The excitation spectrum of a given system can be found using different approaches. In this section, we use for that purpose the field equations of motion, which give the best insight into the physical nature of the various excitation modes. Later on, we will rederive our result for the dispersion relations from the propagator of the pion and photon fields.

As the first step, we shift the magnetic field by the uniform background, $\vek B\to\vek\Bex+\vek B$, and linearize the equations of motion~\eqref{EoM_NR} in the field fluctuations $\phi$, $\vek E$ and $\vek B$,
\begin{equation}
\begin{split}
\ddot\phi-\vek\nabla^2\phi+m_\pi^2\phi&=C\vek\Bex\cdot\vek E,\\
\skal\nabla E&=-C\vek\Bex\cdot\vek\nabla\phi,\\
\vekt\nabla B&=\dot{\vek E}+C\vek\Bex\dot\phi.
\end{split}
\end{equation}
Next, we carry out the Fourier transform to frequency-momentum space by introducing the conjugate variables $\omega$ and $\vek p$, and the corresponding Fourier components of the fields, $\varphi_{\omega,\vek p}$, $\vek e_{\omega,\vek p}$ and $\vek b_{\omega,\vek p}$.\footnote{We will nevertheless drop the subscripts $\omega$ and $\vek p$; since we use different symbols to denote fields in the coordinate and momentum spaces, no confusion can arise.} Upon using the Bianchi identity to eliminate $\vek b$ in favor of $\vek e$, $\vek b=\frac1\omega\vekt pe$, and a brief further manipulation, the set of linearized equations of motion can be cast as
\begin{equation}
\begin{split}
(-\omega^2+\vek p^2+m_\pi^2)\varphi&=C\skal\Bex e,\\
(\omega^2-\vek p^2)\vek e&=C\varphi[\vek p(\skal p\Bex)-\omega^2\vek\Bex].
\end{split}
\end{equation}
This set of equations admits two classes of solutions:
\begin{itemize}
\item One solution with
\begin{equation}
\omega_\gamma(\vek p)=|\vek p|.
\label{displin}
\end{equation}
This solution corresponds to the usual electromagnetic wave, characterized by $\varphi=0$ and $\skal ep=\skal e\Bex=0$. Its electric component is thus linearly polarized in the plane perpendicular to the background field $\vek\Bex$.
\item Two solutions with
\begin{equation}
\omega_\pm^2(\vek p)=\vek p^2+\frac12(m_\pi^2+\Bex^2C^2)\pm\sqrt{\Bex^2C^2\vek p_\perp^2+\frac14(m_\pi^2+\Bex^2C^2)^2},
\label{disprel}
\end{equation}
\end{itemize}
where $\vek p_\perp$ is the component of momentum transverse to $\vek\Bex$. These solutions have nonzero $\varphi$ as well as the electric field, related by
\begin{equation}
\vek e=C\varphi\frac{\vek p(\skal p\Bex)-\omega^2\vek\Bex}{\omega^2-\vek p^2}.
\label{esol}
\end{equation}

The structure of the spectrum is best understood by focusing first on modes propagating along the magnetic field $\vek\Bex$. In this case, the residual $\text{SO}(2)$ rotational symmetry ensures that helicity is a well-defined quantum number. Consequently, the pion and photon modes decouple. The dispersion relation $\omega_-$ becomes degenerate with $\omega_\gamma$. Together, they define the two usual transverse electromagnetic waves, polarized in the plane perpendicular to $\vek\Bex$. Their propagation along the magnetic field $\vek\Bex$ is unaffected by the anomaly as its contribution to the Lagrangian~\eqref{master}, proportional to $\vek E\cdot\vek B$, now vanishes. The presence of two gapless degrees of freedom in the spectrum can thus be considered a consequence of gauge invariance just like in the vacuum of quantum electrodynamics (without background magnetic field).

The remaining excitation propagating along $\vek\Bex$ corresponds to a $\varphi$-wave, accompanied by longitudinal fluctuations of the electric field, as dictated by eq.~\eqref{esol}. The existence of this electric component follows from the fact that a gradient of the pion field induces nonzero electric charge density in presence of the magnetic background, see the equation of motion~\eqref{EoM_NR}. This pion-like mode has a gap,
\begin{equation}
m_\text{eff}\equiv\sqrt{m_\pi^2+\Bex^2C^2},
\label{meff}
\end{equation}
which remains nonzero even in the chiral limit. This can be understood as Schwinger mass generation due to the 1+1-dimensional chiral anomaly, obtained from the 3+1-dimensional anomaly by dimensional reduction in presence of the background magnetic field.\footnote{We are grateful to the referee for suggesting this interpretation of the pion mass in the chiral limit.}

Since the dispersion relations should be continuous functions of momentum, we can conclude from the above discussion that there will be two gapless and one gapped mode for arbitrary direction of propagation. The dispersion relation of the gapless $\omega_-$ mode can be expanded in powers of momentum as
\begin{equation}
\omega_-^2(\vek p)=\vek p_\parallel^2+\frac{m_\pi^2\vek p_\perp^2}{m_\text{eff}^2}+\frac{\Bex^4C^4\vek p_\perp^4}{m_\text{eff}^6}+\mathcal O(\vek p_\perp^6),
\label{disprelexp}
\end{equation}
where $\vek p_\parallel$ is naturally the component of momentum in the direction of $\vek\Bex$. The anisotropy of the dispersion relation becomes maximal in the chiral limit where the $\vek p_\perp^2$ term vanishes and the dispersion relation in the transverse directions becomes quadratic.

At this point, we would like to remark that the observed anomaly-induced mixing of photons with a pseudoscalar (here the neutral pion) is of course not a new concept. It has in particular been known for a long time in the context of axion physics~\cite{Sikivie:1983ip,Sikivie:1985yu,Raffelt:1987im}. However, the physical scale hierarchy is quite different in that context, and the interplay of the extremely weak photon-axion interaction with the coherence of the photon beam results in a photon-axion \emph{conversion} rather than full mixing; see also ref.~\cite{Akhmedov:2009rb} for a related discussion within neutrino physics. To the best of our knowledge, the actual mixing problem was first analyzed in ref.~\cite{Maiani:1986md}, and solved therein for the special case of propagation in the plane transverse to $\vek\Bex$. Our dispersion relations~\eqref{disprel}, valid for an arbitrary direction of momentum, are a generalization thereof.


\subsection{Alternative derivation}
\label{subsec:alternative}

It is instructive to rederive the dispersion relations~\eqref{disprel} directly from the covariant equations of motion~\eqref{EoM_rel}; this allows us to introduce already now some notation that we will make use of later in the calculation of the pion decay with. To make contact with that follow-up quantum-field-theoretic calculation, we keep the gauge-fixing term in the equation of motion for $A^\mu$, but select the Feynman gauge, $\xi=1$, in which the equation takes a particularly simple form.\footnote{The dispersion relations of the physical modes are of course independent of such a choice of gauge. However, we will later on also need the photon polarization vector, which does depend on the gauge.}

Denoting the Fourier component of $A^\mu$ as $a^\mu_p$, the linearized version of the equations of motion~\eqref{EoM_rel} can then be written compactly as
\begin{equation}
\begin{split}
(p^2-m_\pi^2)\varphi-\imag n\cdot a&=0,\\
p^2a^\mu-\imag\varphi n^\mu&=0,
\end{split}
\label{EoM_relativistic}
\end{equation}
where
\begin{equation}
n_p^\mu\equiv\frac C2\eps^{\mu\nu\alpha\beta}p_\nu F^\text{ex}_{\alpha\beta}.
\label{nvector}
\end{equation}
The most straightforward way to solve these equations is to contract the second of them with $n_\mu$ and thereby eliminate $\varphi$ in terms of $n\cdot a$. This leads to two classes of solutions, which are in a one-to-one correspondence with the solutions found above using the nonrelativistic equations of motion~\eqref{EoM_NR}:
\begin{itemize}
\item One solution with $n\cdot a=0$, for which $\varphi=0$ and the momentum satisfies $p^2=0$. This is the relativistic linearly polarized photon.
\item Two solutions with $n\cdot a\neq0$, for which momentum satisfies the covariant condition
\begin{equation}
p^2(p^2-m_\pi^2)+n^2=0.
\label{polecond}
\end{equation}
\end{itemize}
Given that $n^2=\Bex^2C^2(\vek p_\parallel^2-\omega^2)$, it is easy to check that this reproduces the dispersion relations $\omega_\pm(\vek p)$ in eq.~\eqref{disprel}.


\section{Pressure at one loop}
\label{sec:pressure}

Pressure is one of the most important observables relevant for both model calculations and lattice simulations of the QCD phase diagram. In this section, we evaluate the \emph{thermal} part of the pressure of our system; in other words, we do not investigate the effect of the magnetic field on the zero-temperature pressure of QCD. At the leading, one-loop order of the loop expansion, the pressure due to thermal excitations is given simply by
\begin{equation}
P=-T\sum_{i=1}^3\int\frac{\dd^3\!\vek p}{(2\pi)^3}\log\bigl[1-e^{-\beta\omega_i(\vek p)}\bigr],
\label{Pdef}
\end{equation}
where the sum runs over all physical excitations in the system and $\omega_i(\vek p)$ are their respective dispersion relations, here given by eqs.~\eqref{displin} and~\eqref{disprel}. This expression describes a free gas of noninteracting quasiparticles with dispersion relations fixed to their zero-temperature values. The feedback of the thermal medium into the spectrum only enters the pressure at the two-loop order through the quasiparticle interactions.

Following our notation for the individual dispersion relations, we split the pressure into the contributions of the individual modes, $P=P_\gamma+P_++P_-$. The contribution of the relativistic photon mode with the linear dispersion relation is trivial to evaluate and amounts to the usual expression,
\begin{equation}
P_\gamma=\frac{\pi^2T^4}{90}.
\label{Pgamma}
\end{equation}
As to the other two modes, we will in the following focus on the gapless mode $\omega_-$ and discard the contribution $P_+$. This is well justified at temperatures $T\ll m_\text{eff}$ where this contribution is exponentially suppressed due to the gap of the $\omega_+$ mode. By using spherical coordinates and integrating by parts with respect to the radial momentum, the momentum integral in eq.~\eqref{Pdef} can be rewritten in the dimensionless form
\begin{equation}
P_-=\frac{T^4}{12\pi^2}\int_0^\pi\dd\!\theta\sin\theta\int_0^\infty\dd\!x\,\frac{x^3\frac{\de\tilde\omega_-}{\de x}}{e^{\tilde\omega_-}-1},
\label{dimlessint}
\end{equation}
where the dimensionless dispersion relation $\tilde\omega_-$ is defined by
\begin{equation}
\tilde\omega_-(x,\theta)\equiv\frac{\omega_-(\vek p)}T=\frac1\tau\biggl(\frac12+x^2\tau^2-\sqrt{\frac14+x^2\tau^2\sin^2\alpha\sin^2\theta}\biggr)^{1/2},
\end{equation}
$\theta$ is the angle between the momentum vector and the direction of the background field $\vek\Bex$, and the parameters $\alpha$ and $\tau$ are defined by
\begin{equation}
\cos\alpha\equiv\frac{m_\pi}{m_\text{eff}}=\frac{m_\pi}{\sqrt{m_\pi^2+\Bex^2C^2}},\qquad
\tau\equiv\frac T{m_\text{eff}}.
\label{defalpha}
\end{equation}
While we cannot evaluate the thermal integral involving the dimensionless dispersion relation $\tilde\omega_-$ in a closed form, eq.~\eqref{dimlessint} is suitable for an expansion in powers of $\tau$, or equivalently in powers of temperature. (Recall that we assume the temperature to be small compared to $m_\text{eff}$, and thus $\tau$ to be much smaller than one.) Both the radial and the angular integral is straightforward to carry out upon such expansion, thus leading to the series expansion of the pressure $P_-$,
\begin{equation}
P_-=\frac{\pi^2T^4}{90}\left[1+\frac{\Bex^2C^2}{m_\pi^2}-\frac{32\pi^2}{21}\frac{\Bex^4C^4T^2}{m_\pi^6}+\frac{384\pi^4}{35}\frac{\Bex^6C^6T^4}{m_\pi^{10}}+\mathcal O(T^6)\dotsb\right],
\label{Pmassive}
\end{equation}
with the expansion parameter $\Bex^2C^2T^2/m_\pi^4$. Interestingly, the expansion is simultaneously an expansion in powers of $\Bex$: the pion mass in the denominators is the ``bare'' mass $m_\pi$ rather than the effective mass in the magnetic field, $m_\text{eff}$. It is therefore mandatory to inspect separately what happens in the chiral limit where $m_\pi$ goes to zero.

The crucial observation is that in this limit, the dispersion relation in the transverse directions becomes quadratic in momentum, see eq.~\eqref{disprelexp}. Obviously, we have to treat the transverse and longitudinal directions separately, and it is therefore most natural to use cylindrical coordinates to carry out the momentum integration. In contrast to eq.~\eqref{dimlessint}, the pressure of the gapless mode can now be rewritten in the dimensionless form
\begin{equation}
P_-=-\frac{\Bex CT^3}{8\pi^2}\int_0^\infty\dd\!x\dd\!y\log\bigl(1-e^{-\tilde{\tilde\omega}_-}\bigr),
\end{equation}
where $x$ is the dimensionless longitudinal momentum, $y$ is likewise the dimensionless transverse radial momentum \emph{squared}, and
\begin{equation}
\tilde{\tilde\omega}_-(x,y)\equiv\frac{\omega_-(\vek p)}T=\frac1\tau\biggl(\frac12+x^2\tau^2+y\tau-\sqrt{\frac14+y\tau}\biggr)^{1/2}.
\end{equation}
Remarkably, the resulting two-dimensional integral can be evaluated in a closed form, for instance by introducing a new variable $z$ via the substitution
\begin{equation}
z^2\tau^2\equiv\frac12+y\tau-\sqrt{\frac14+y\tau}=\frac14(\sqrt{1+4y\tau}-1)^2,
\end{equation}
and subsequently using polar coordinates in the $xz$ plane. The final result for the pressure due to the gapless mode $\omega_-$ then reads
\begin{equation}
P_-\Bigr|_{m_\pi=0}=\frac{\zeta(3)}{16\pi}\Bex CT^3+\frac{\pi^2T^4}{180}.
\label{Pchiral}
\end{equation}
As long as the contribution of the gapped mode $\omega_+$ can be neglected, the full one-loop pressure of the system is given without further approximations by the closed expression
\begin{equation}
P\Bigr|_{m_\pi=0}=P_\gamma+P_-\Bigr|_{m_\pi=0}=\frac{\zeta(3)}{16\pi}\Bex CT^3+\frac{\pi^2T^4}{60}.
\end{equation}
One can even say that this compact expression represents the complete asymptotic series expansion of the pressure at low temperatures. Namely, the contribution of the $\omega_+$ mode is suppressed by the non-analytic Boltzmann factor $e^{-m_\text{eff}/T}$, as a consequence of which all of its Taylor coefficients at $T=0$ vanish.

We can conclude that the mixing of neutral pions with dynamical photons dramatically changes the thermodynamics of QCD at low temperatures and in strong magnetic fields: instead of the usual black-body scaling with $T^4$, the pressure at temperatures $T\ll\Bex C$ is dominated by a term that scales as $T^3\Bex/f_\pi$. It is interesting to contrast this to the $T^{5/2}(\Bex/f_\pi)^{3/2}$ scaling at \emph{nonzero} baryon chemical potential, found in ref.~\cite{Brauner:2017mui}.


\section{Anomalous pion decay}
\label{sec:decay}

The presence of the strong background magnetic field affects also other physical observables than those relevant for equilibrium thermodynamics. In this section, we will focus on the decay properties of the neutral pion, bearing in mind that in the vacuum, the dominant, two-photon decay of the pion is one of the hallmarks of the chiral anomaly (see ref.~\cite{Ioffe:2007eg} for a recent precision calculation of the vacuum anomalous pion decay rate). The magnetic field affects the neutral pion decay in several ways. First, it contributes to the pion mass through loop corrections~\cite{Agasian:2001ym,Andersen:2012dz,Andersen:2012zc}. Second, it affects, likewise through loop corrections, the pion decay constant $f_\pi$~\cite{Simonov:2015xta}. Finally, it may open phase space for new decay processes, or increase the branching ratio of processes that in the vacuum are negligible compared to the two-photon decay~\cite{Bar:2004bw,Hattori:2013cra}. Our discussion below focuses on the consequences of the chiral anomaly in a background magnetic field for the two-photon pion decay rate at tree level, in particular on the effects of the kinematic mixing of pions and photons. Non-anomalous loop corrections due to the magnetic field are not included here; their significance is discussed briefly in the concluding section~\ref{sec:summary}.

For the sake of simplicity, we shall in this section refer to the $\omega_+$ mode as the ``pion'', and denote the corresponding one-particle state with momentum $\vek p$ as $\ket{\pi,\vek p}$. This is natural for this state smoothly interpolates to the vacuum neutral pion in the limit of vanishing magnetic field, $\Bex\to0$. Likewise, we shall refer to the other two states as ``photons'', using the following notation:
\begin{itemize}
\item$\ket{\Gamma,\vek p}$ for the ``nonrelativistic'' photon with the dispersion relation $\omega_-(\vek p)$.
\item$\ket{\gamma,\vek p}$ for the ``relativistic'' photon with the dispersion relation $\omega_\gamma(\vek p)$.
\end{itemize}
Without further mentioning it explicitly, we shall use the Feynman gauge in which $\xi=1$.

The two-photon decay of the neutral pion in principle includes three different processes: $\pi\to\gamma\gamma$, $\pi\to\Gamma\Gamma$ and $\pi\to\gamma\Gamma$; this corresponds to the sum over polarizations of the photons in the final state. However, only the last process, $\pi\to\gamma\Gamma$, is actually allowed. The reason for this is that QCD in a background magnetic field possesses a discrete symmetry, which can be thought of as modified parity, under which $\gamma$ is even whereas $\pi$ and $\Gamma$ are odd~\cite{Raffelt:1987im}. Below, we therefore calculate the decay rate for the $\pi\to\gamma\Gamma$ process alone and compare it to the vacuum decay rate of the neutral pion. (We have checked by an explicit calculation that the decay rates for the $\pi\to\gamma\gamma$ and $\pi\to\Gamma\Gamma$ processes vanish.)


\subsection{Relativistic photon polarization vector}
\label{subsec:polarization}

In order to determine the probability rate for any decay or scattering process, it is mandatory to understand the properties of the asymptotic one-particles states. We already know the dispersion relations of all the one-particle states in our system. However, the calculation of decay rates or scattering cross-sections also requires the knowledge of the corresponding wave functions, or in other words of how the one-particle states couple to the elementary fields in our EFT. We start here with the discussion of the relativistic photon case, which is the most subtle of the three one-particle states as it is affected by gauge ambiguities.

Recall the field equations of motion in the relativistic notation and the Feynman gauge, eq.~\eqref{EoM_relativistic}. In the second quantization, the solutions to these equations of motion enter the plane-wave expansion of the pion and electromagnetic fields. In particular, the relativistic photon mode has $\varphi=0$, and thus only appears in the expansion of the field $A_\mu$ via
\begin{equation}
A_\mu(x)\supset\int\frac{\dd^3\!\vek p}{(2\pi)^{3/2}\sqrt{2|\vek p|}}\left[\eps_\mu(\vek p)a_{\vek p}e^{-\imag p\cdot x}+\eps^*_{\mu}(\vek p)\he a_{\vek p}e^{\imag p\cdot x}\right],
\end{equation}
where $\he a_{\vek p}$ is the creation operator of the corresponding one-particle state, $\ket{\gamma,\vek p}=\he a_{\vek p}\ket0$, $a_{\vek p}$ is the associated annihilation operator, and $\eps_\mu(\vek p)$ is the polarization vector of the one-particle state. The four-momentum $p$ in the integral is on-shell, that is, $p^0=\omega_\gamma(\vek p)=|\vek p|$. The polarization vector $\eps_{\mu}$ must satisfy the following conditions:
\begin{itemize}
\item The transversality constraint $n\cdot\eps=0$, following from the property $n\cdot a=0$ of the relativistic photon solution to the classical equations of motion~\eqref{EoM_relativistic}.
\item The Feynman gauge constraint $p\cdot\eps=0$. Note that upon using the on-shell condition $p^2=0$, this constraint is invariant under the gauge transformation $\eps^\mu\to\eps^\mu+\lambda p^\mu$ with arbitrary complex $\lambda$; the same gauge invariance property must apply to all physical observables.
\end{itemize}
The freedom to shift $\eps^\mu$ without affecting physical observables can be used to set $\eps^0=0$ without loss of generality. We thus have altogether three linear constraints on $\eps^\mu$,
\begin{equation}
\eps^0=0,\qquad
n\cdot\eps=0,\qquad
p\cdot\eps=0,
\label{polvector}
\end{equation}
which in four spacetime dimensions determine $\eps^\mu$ uniquely up to an overall factor. We note that given the explicit expression for the vector $n^\mu$, eq.~\eqref{nvector}, it is easy to find an explicit solution to these constraints in a covariant form,
\begin{equation}
\eps^\mu(\vek p)\propto F^{\mu\nu}_\text{ex}p_\nu.
\label{polvectorexplicit}
\end{equation}
The overall normalization will be fixed using a different argument in the next subsection.


\subsection{Vacuum transition amplitudes}
\label{subsec:amplitudes}

In presence of field mixing, the one-to-one correspondence between elementary fields and asymptotic one-particle states is lost. Physical scattering amplitudes then have to be extracted from the Green's functions of the fields using the Lehmann-Symanzik-Zimmermann reduction formula. To that end, we need the vacuum transition amplitudes $\bra0\chi_i(0)\ket{n,\vek p}$, where $\chi_i(0)$ runs over all elementary fields of the theory and $n$ over all one-particle states. Consider the matrix propagator of the elementary fields, that is, the time-ordered two-point Green's function, defined by $\De_{ij}(x-y)\equiv-\imag\bra0T\{\chi_i(x)\chi_j(y)\}\ket0$. By inserting the partition of unity in terms of the eigenstates of the Hamiltonian, one arrives at the K\"all\'en-Lehmann spectral representation of the propagator in its nonrelativistic form~\cite{Brauner:2006xm},
\begin{equation}
\De_{ij}(\omega,\vek p)=(2\pi)^3\sum_n\left[\frac{\bra0\chi_i(0)\ket{n,\vek p}\bra{n,\vek p}\chi_j(0)\ket0}{\omega-\omega_n(\vek p)+\imag\varepsilon}-\frac{\bra0\chi_j(0)\ket{n,-\vek p}\bra{n,-\vek p}\chi_i(0)\ket0}{\omega+\omega_n(\vek p)-\imag\varepsilon}\right],
\label{kallen}
\end{equation}
where $\omega_n(\vek p)$ denotes the energy of the Hamiltonian eigenstate $\ket{n,\vek p}$. These eigenstates are assumed to be normalized according to $\langle n,\vek p|m,\vek q\rangle=\delta_{mn}\delta^3(\vek p-\vek q)$. For one-particle states, $n$ takes discrete values and the propagator correspondingly has simple poles at energies given by the dispersion relations of the one-particle states. For multi-particle states, on the other hand, the label $n$ is continuous, resulting in a branch cut in the propagator. Equation~\eqref{kallen} tells us that the vacuum transition amplitudes connecting elementary fields to a given one-particle state can be extracted from the residuum of the associated pole in the propagator.

The \emph{inverse} propagator can be directly read off the Lagrangian~\eqref{master}. It is a $5\times5$ matrix and upon Fourier transform to momentum space and setting $\xi=1$ takes the form
\begin{equation}
\renewcommand{\arraystretch}{1.3}
\De^{-1}_{\mu\nu}(p)=
\left(\begin{array}{c|c}
p^2-m_\pi^2 & -\imag n_\nu\\
\hline
+\imag n_\mu & -g_{\mu\nu}p^2
\end{array}\right).
\end{equation}
Note that we used a somewhat unusual notation wherein the first row and column corresponds to the $\phi$ field, and the remaining four rows and columns correspond to the $A_\mu$ field and carry the index $\mu$ and $\nu$, respectively. Inverting this $5\times5$ matrix and using the fact that $n\cdot p=0$ leads to an expression for the propagator,
\begin{equation}
\renewcommand{\arraystretch}{2}
\De_{\mu\nu}(p)=
\left(\begin{array}{c|c}
\displaystyle\frac{p^2}{p^2(p^2-m_\pi^2)+n^2} & \displaystyle-\frac{\imag n_\nu}{p^2(p^2-m_\pi^2)+n^2}\\[1ex]
\hline
\displaystyle+\frac{\imag n_\mu}{p^2(p^2-m_\pi^2)+n^2} & \displaystyle-\frac{g_{\mu\nu}}{p^2}+\frac{n_\mu n_\nu}{p^2[p^2(p^2-m_\pi^2)+n^2]}.
\end{array}\right)
\label{propagator}
\end{equation}
The pole structure of the propagator naturally reproduces our previous results for the quasiparticle dispersion relations, see e.g.~eq.~\eqref{polecond}.

Let us first have a look at the poles corresponding to the two nonrelativistic modes, $\pi$ and $\Gamma$. Comparing eq.~\eqref{propagator} to the general spectral representation of the propagator~\eqref{kallen}, we find that up to an overall phase,
\begin{equation}
\begin{split}
\bra0\phi(0)\ket{\pm,\vek p}&=\pm\frac1{(2\pi)^{3/2}}\sqrt{\frac{\Delta_{\vek p}\pm m_\text{eff}^2}{4\omega_\pm(\vek p)\Delta_{\vek p}}},\\
\bra0A_\mu(0)\ket{\pm,\vek p}&=\frac\imag{(2\pi)^{3/2}}\frac{n^p_\mu}{\sqrt{\omega_\pm(\vek p)\Delta_{\vek p}(\Delta_{\vek p}\pm m_\text{eff}^2)}},
\end{split}
\label{transamp1}
\end{equation}
where we defined
\begin{equation}
\Delta_{\vek p}\equiv\omega_+^2(\vek p)-\omega_-^2(\vek p)=\sqrt{4\Bex^2C^2\vek p_\perp^2+m_\text{eff}^4},
\end{equation}
and temporarily denoted the states $\ket{\pi,\vek p}$ and $\ket{\Gamma,\vek p}$ as $\ket{+,\vek p}$ and $\ket{-,\vek p}$, respectively, in order to keep the expressions compact.

The coupling of the fields to the relativistic photon state $\gamma$ can likewise be extracted by looking at the pole at $p^2=0$ in the propagator~\eqref{propagator}. Here the situation is, however, complicated by the fact that the propagator in covariant gauges includes contributions from states in the unphysical sector of the Hilbert space. This is also easily seen from the fact that the lower-right $4\times4$ corner of the propagator is proportional to $-g_{\mu\nu}+n_\mu n_\nu/n^2$ at $p^2=0$, which up to a sign is a projector to the three-dimensional space of vectors orthogonal to $n_\mu$, although there is only one physical state with the dispersion relation given by $p^2=0$. The way out is to notice that by the argument of the preceding subsection, $\bra0A_\mu(0)\ket{\gamma,\vek p}$ must be proportional to $\eps_\mu(\vek p)$. This fixes the one-dimensional subspace, corresponding to the physical photon polarization. The overall normalization of the vacuum transition amplitude can then be fixed using eqs.~\eqref{kallen} and~\eqref{propagator}, leading to
\begin{equation}
\begin{split}
\bra0\phi(0)\ket{\gamma,\vek p}&=0,\\
\bra0A_\mu(0)\ket{\gamma,\vek p}&=\frac1{(2\pi)^{3/2}\sqrt{2|\vek p|}}\frac{\eps_\mu(\vek p)}{\sqrt{-\eps(\vek p)^2}}.
\end{split}
\label{transamp2}
\end{equation}
Equations~\eqref{transamp1} and~\eqref{transamp2} form the basis for our calculation of the pion decay rate below.


\begin{figure}
$$
\includegraphics[scale=1]{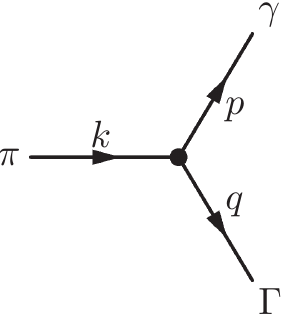}
$$
\caption{Kinematics of the $\pi\to\gamma\Gamma$ process, indicating our notation for the momenta of the three states; the arrows indicate the flow of momentum. We only consider the decay of pion at rest here, in which case $k^\mu=(m_\text{eff},\vek0)$.}
\label{fig:kinematics}
\end{figure}

\subsection{Kinematics}
\label{subsec:kinematics}

Before we move on to the calculation of the amplitude for the pion decay, we first discuss the kinematics of the $\pi\to\gamma\Gamma$ process. Since Lorentz invariance is explicitly broken by the presence of the background magnetic field, the decay rate in principle has to be evaluated as a function of velocity or momentum. We limit our attention for the sake of simplicity to decay of a pion \emph{at rest}, since this assumption, as we will see, leads to a simple, semi-analytic expression for the decay rate. We expect the result to give a reasonable approximation also for a nonzero velocity of the pion provided that it is much smaller than the speed of light.

The kinematics of the decay process is displayed in figure~\ref{fig:kinematics}. Momentum conservation in the rest frame of the pion leads trivially to $\vek p+\vek q=\vek0$. Imposing the energy conservation condition, $m_\text{eff}=|\vek p|+\omega_-(\vek q)$, then gives the magnitude of momentum of the particles in the final state as a function of the angle $\theta$ with respect to the magnetic field $\vek\Bex$,
\begin{equation}
|\vek p|=\frac{m_\text{eff}/2}{1-\bigl(\frac{\Bex C}{2m_\text{eff}}\bigr)^2\sin^2\theta}=\frac{m_\text{eff}}{2-\frac12\sin^2\alpha\sin^2\theta},
\label{momcons}
\end{equation}
using the notation introduced in eq.~\eqref{defalpha}.

In the next subsection, we will calculate the amplitude $\Me$ for the $\pi\to\gamma\Gamma$ decay at tree level, including the corresponding vacuum transition amplitudes that couple fields to one-particle states. The differential decay rate for the process then reads
\begin{equation}
\dd\!\Gamma=(2\pi)^4\delta^4(k-p-q)\times(2\pi)^3|\Me|^2\dd^3\!\vek p\dd^3\!\vek q.
\end{equation}
The $\delta$-function, imposing energy and momentum conservation, reduces the phase space integration to an angular integration over directions of momentum $\vek p$,
\begin{equation}
\Gamma=(2\pi)^7\int\dd^3\!\vek p\,|\Me|^2\delta\bigl(m_\text{eff}-|\vek p|-\omega_-(\vek p)\bigr)=(2\pi)^7\int\dd\!\Omega\,|\Me|^2\frac{|\vek p|^2}{1+\frac{\dd\!\omega_-}{\dd\!|\vek p|}},
\label{Gamma}
\end{equation}
where $|\vek p|$ is determined by eq.~\eqref{momcons}.


\begin{figure}
$$
\parbox{25mm}{%
\includegraphics[scale=1]{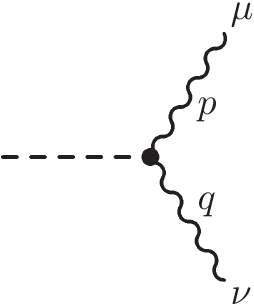}}=-\imag C\eps^{\mu\nu\alpha\beta}p_\alpha q_\beta.
$$
\caption{Feynman rule for the interaction between pion and electromagnetic fields, following from the Lagrangian~\eqref{master}. The dashed line stands for the $\phi$ field, whereas the wavy lines for the $A$ field. The momenta $p$ and $q$ flow \emph{out} of the vertex.}
\label{fig:Frule}
\end{figure}

\subsection{Decay rate}
\label{subsec:rate}

The decay of the pion into a photon pair is driven by a single interaction vertex in the Lagrangian~\eqref{master}, containing both fields. The corresponding Feynman rule is shown in figure~\ref{fig:Frule}. The calculation of the decay amplitude is, however, complicated by the kinematic mixing between the fields. As a consequence, the $\phi$ field in the interaction operator can couple to both the $\pi$ in the initial state and the $\Gamma$ in the final state, and so can one of the $A$ fields. The other $A$ must necessarily couple to the $\gamma$ in the final state, since $\bra0\phi\ket{\gamma,\vek p}=0$. The decay amplitude therefore consists of two contributions,
\begin{equation}
\begin{split}
-\imag\Me={}&-\imag C\eps^{\mu\nu\alpha\beta}p_\alpha q_\beta\bra{\gamma,\vek p}A_\mu\ket0\bra{\Gamma,\vek q}A_\nu\ket0\bra0\phi\ket{\pi,\vek k}\\
&-\imag C\eps^{\mu\nu\alpha\beta}p_\alpha(-k)_\beta\bra{\gamma,\vek p}A_\mu\ket0\bra{\Gamma,\vek q}\phi\ket0\bra0A_\nu\ket{\pi,\vek k}.
\end{split}
\end{equation}
Next one inserts the vacuum transition amplitudes from eqs.~\eqref{transamp1} and~\eqref{transamp2} and takes the squared absolute value of the amplitude. What follows is a rather lengthy calculation, including manipulation of products of Levi-Civita tensors and kinematic properties of the momenta $k$, $p$, $q$, the vectors $n_k^\mu$ and $n_q^\mu$, and the polarization vector $\eps^\mu_p$. At the end of the calculation, a very compact result surfaces,
\begin{equation}
|\Me|^2=\frac{C^2}{(2\pi)^9}\frac{\Delta_{\vek p}+m_\text{eff}^2}{16m_\text{eff}|\vek p|\omega_-(\vek p)\Delta_{\vek p}}\left[m_\text{eff}\omega_-(\vek p)-\frac12(\Delta_{\vek p}-m_\text{eff}^2)\right]^2.
\end{equation}
Note that in the limit $\Bex\to0$, this expression further simplifies to $C^2m_\pi/[8(2\pi)^9]$, and upon trivial angular integration following eq.~\eqref{Gamma} gives
\begin{equation}
\Gamma_\text{vac}=\frac{C^2m_\pi^3}{64\pi},
\label{Gammavac}
\end{equation}
which agrees with the well-known expression for the neutral pion decay rate in the vacuum.  For nonzero values of the background field, one simply has to insert the squared amplitude in eq.~\eqref{Gamma} and simplify the result. It is natural to normalize the decay rate by its value $\Gamma_\text{vac}$ in zero magnetic field. Upon some manipulation, we thus obtain the final result for the magnetic field dependence of the decay rate of the neutral pion,
\begin{equation}
\frac{\Gamma(\Bex)}{\Gamma_\text{vac}}=\frac1{\cos^3\alpha}\biggl\langle\frac{(1-\sin^2\alpha\sin^2\theta)^2}{(1-\frac14\sin^2\alpha\sin^2\theta)^4}\biggr\rangle_\Omega,
\label{finalresult}
\end{equation}
where the angular brackets indicate angular averaging over the full solid angle corresponding to the variable $\theta$, and the angle $\alpha$ is defined in eq.~\eqref{defalpha}.\footnote{In fact, the angular averaging indicated in eq.~\eqref{finalresult} can be carried out analytically in a closed form. However, the resulting expression is rather cumbersome, and we therefore prefer the simple form of eq.~\eqref{finalresult}; the angular average can, if needed, be done numerically with no effort.}

\begin{figure}
\begin{center}
\includegraphics[scale=1.2]{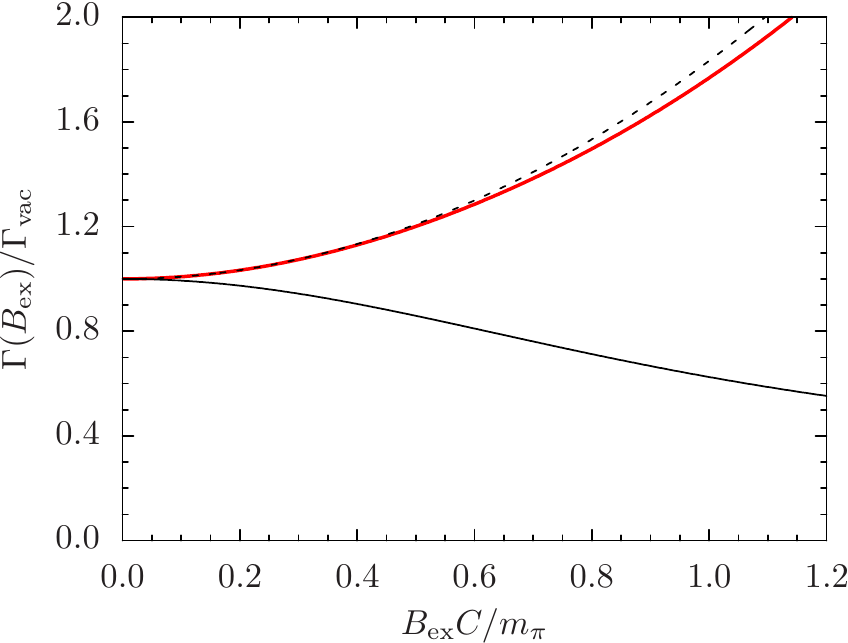}
\end{center}
\caption{Thick red line: dependence of the pion decay rate on the scaled magnetic field, given by eq.~\eqref{finalresult}. The decay rate is normalized to unity in zero magnetic field; the absolute value of the decay rate in the vacuum is determined by eq.~\eqref{Gammavac}. Dashed black line: polynomial approximation for the decay rate given by the first two terms in eq.~\eqref{alphaexp}. Solid black line: the decay rate without the $1/\cos^3\alpha$ prefactor, that is, just the anisotropy factor in eq.~\eqref{finalresult}. While all the numerical values only depend on the dimensionless ratio $\Bex C/m_\pi$, the range on the horizontal axis was chosen so that, for physical values of $m_\pi$ and $f_\pi$, its upper limit corresponds to $\Bex\approx10^{20}\text{ G}$.}
\label{fig:Gamma}
\end{figure}

The dependence of the pion decay rate on the magnetic field is shown numerically in figure~\ref{fig:Gamma}. The increase of the decay rate with increasing $\Bex$ is, in fact, a result of a competition of two effects. First, the phase space for the decay products is increased due to the increase of the pion mass from $m_\pi$ to $m_\text{eff}$; this corresponds to the leading $1/\cos^3\alpha$ factor in eq.~\eqref{finalresult}. Second, the anisotropic kinematics reduces somewhat the result by the angular average factor in eq.~\eqref{finalresult}. Obviously, the anomalous contribution to the pion mass plays a dominant role for the pion decay.

Apart from the full numerical result, analytical approximations for the decay rate may also be of some interest. Given eq.~\eqref{finalresult}, it is straightforward to obtain power expansions for the decay rate in both weak and strong magnetic fields,
\begin{align}
\label{alphaexp}
\frac{\Gamma(\Bex)}{\Gamma_\text{vac}}&=1+\frac56\tan^2\alpha-\frac{19}{120}\tan^4\alpha+\frac{251}{1680}\tan^6\alpha+\mathcal O(\tan^8\alpha),\\
\frac{\Gamma(\Bex)}{\Gamma_\text{vac}}&=\left(\frac{8\pi}{9\sqrt3}-\frac43\right)\tan^3\alpha+\left(\frac{52\pi}{27\sqrt3}-\frac83\right)\tan\alpha+\mathcal O(\tan^{-1}\alpha).
\end{align}
The latter expansion is particularly relevant for the chiral limit where only the leading term survives and we get a closed expression for the decay rate, this time in absolute units,\footnote{In the chiral limit, the phase space for the pion decay is closed in the vacuum, see also eq.~\eqref{Gammavac}. The chiral anomaly then provides two key ingredients that make the pion decay possible in background magnetic fields: both the interaction with photons and the phase space by giving the pion a nonzero mass.}
\begin{equation}
\Gamma(\Bex)\Bigr|_{m_\pi=0}=\frac{\Bex^3C^5}{64\pi}\left(\frac{8\pi}{9\sqrt3}-\frac43\right).
\label{chirallimit}
\end{equation}


\section{Summary and discussion}
\label{sec:summary}

In this paper, we have analyzed the low-temperature thermodynamics of QCD in strong magnetic fields. This is dominated by neutral pions and photons since the charged pions acquire a large gap due to Landau level quantization. We showed that the chiral anomaly leads in presence of the background magnetic field to mixing of neutral pions and photons, and worked out the consequences of this mixing for physical observables.

Our first main result is an expression for pressure of the system in the leading, one-loop approximation, see eqs.~\eqref{Pmassive} and~\eqref{Pchiral}. The softening of the dispersion relation of one of the photon polarizations due to the mixing leads to an enhancement of pressure at low temperatures, which is most dramatic in the chiral limit, where the leading contribution to pressure scales as $\Bex T^3/f_\pi$.

Our second main result is a formula for the dependence of the neutral pion decay rate on the magnetic field, eq.~\eqref{finalresult}. The effect of the magnetic field is again most dramatic in the chiral limit. The $\Bex^3/f_\pi^5$ scaling, displayed in eq.~\eqref{chirallimit}, is actually easy to understand. Namely, at tree level, the decay rate must consist of a factor $C^2$ from the interaction vertex times a kinematic function of the product $\Bex C$, entering the dispersion relations of the pion and photon. Dimensional analysis then fixes the powers of $\Bex$ and $f_\pi$ in the final result. The nontrivial part of our result therefore is the numerical factor in eq.~\eqref{chirallimit}.

We would now like to discuss some of the assumptions and approximations underlying our analysis in the form of a set of concluding remarks.

\begin{figure}
\begin{center}
\includegraphics[scale=1.2]{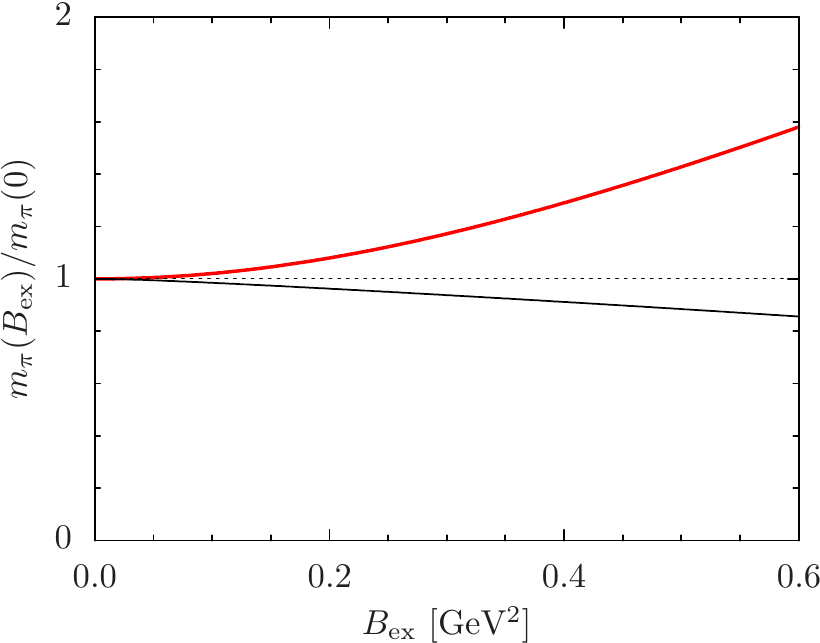}
\end{center}
\caption{Dependence of the \emph{neutral} pion mass on the external magnetic field in units of the vacuum pion mass. Thick red line: the tree-level anomalous mass, given by eq.~\eqref{meff}. Thin black line: the non-anomalous, one-loop mass according to refs.~\cite{Andersen:2012dz,Andersen:2012zc}. The dashed black line is added just to guide the eye. The numerical results were obtained using the physical values $f_\pi\approx92\text{ MeV}$ and $m_\pi\approx135\text{ MeV}$; the anomaly contribution dominates over the non-anomalous loop correction for $\Bex\gtrsim0.1\text{ GeV}^2$. The upper end of the displayed range for $\Bex$ corresponds to $10^{20}\text{ G}$.}
\label{fig:1loop}
\end{figure}

First of all, we worked for the sake of simplicity strictly at the tree level, that is, we neglected one-loop corrections to the vacuum pion mass in presence of a magnetic field, see e.g.~ref.~\cite{Andersen:2012zc}. While those are necessarily proportional to the pion mass itself in accord with the chiral symmetry, the anomaly makes the pion massive even in the chiral limit. Hence its effect will certainly be dominant in sufficiently strong magnetic fields, or sufficiently close to the chiral limit. At the physical point, the anomaly becomes dominant in moderate fields, $B\gtrsim0.1\text{ GeV}^2$, see figure~\ref{fig:1loop}.

\begin{figure}
\begin{center}
\includegraphics[scale=1.2]{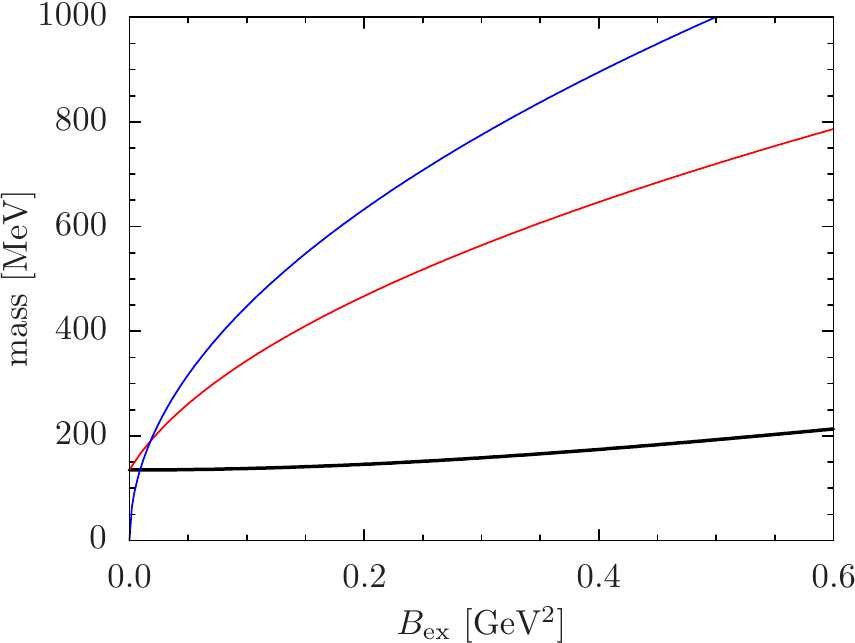}
\end{center}
\caption{Comparison of various energy scales in our system as a function of the external magnetic field. The thick black line corresponds to the anomalous neutral pion mass, eq.~\eqref{meff}. The red line stands for the charged pion mass, lifted by the magnetic field as a consequence of Landau level quantization. Finally, the blue line indicates the threshold for the dilepton ($e^+e^-$) decay of the neutral pion in presence of the external magnetic field. There is clearly a range of magnetic fields in which the pion spectrum features scale separation, that is, the neutral pion is considerably lighter than the charged pion. For illustration, demanding that the ratio of the pion masses is at most $1/2$ requires that $0.06\text{ GeV}^2\lesssim\Bex\lesssim3.2\text{ GeV}^2$, or $10^{19}\text{ G}\lesssim\Bex\lesssim5\times10^{20}\text{ G}$. The numerical results were obtained using the vacuum mass $m_\pi\approx135\text{ MeV}$.}
\label{fig:masses}
\end{figure}

Second, we only kept neutral pions in our EFT, which assumes that there is sufficient scale separation between the neutral and charged pion sectors. The effective mass of the charged pion in the background magnetic field is determined by the standard Landau level problem, $m_{\pi^\pm}(\Bex)=\sqrt{m_\pi^2+\Bex}$. The graphical illustration of the numerical values of the neutral and charged pion masses as a function of the magnetic field in figure~\ref{fig:masses}, makes it clear that there is a large range of magnetic fields in which the requirement of scale separation is satisfied.

In ref.~\cite{Hattori:2013cra} it was argued that in sufficiently strong magnetic fields, the dilepton decay $\pi^0\to e^+e^-$ will be the dominant decay channel for neutral pion. However, it seems that these authors only included the effect of the magnetic field on the amplitude for such decay, not its consequences for the phase space of the decay products. Namely, the energy levels of the electron-positron pair also undergo Landau level quantization. By spin conservation, the decay is only possible into one gapless and one gapped fermion. Hence the threshold energy for the $\pi^0\to e^+e^-$ decay channel to be open altogether equals $m_e+\sqrt{m_e^2+2\Bex}$, where $m_e$ is the vacuum electron mass. The position of the threshold is indicated by the blue line in figure~\ref{fig:masses}, which makes it clear that the dilepton channel is actually closed in most of the range displayed therein, except for fields below ca $10^{18}\text{ G}$.

Finally, we neglected nonlinear effects within electrodynamics, which are induced by loop corrections and in presence of the background magnetic field lead to vacuum birefringence and photon splitting~\cite{Adler:1970gg,BialynickaBirula:1970vy,Adler:1971wn}. While these effects in general modify the photon polarization tensor, and thus its propagation in the background field, they will not affect the main qualitative conclusions of our paper.


\acknowledgments
T.B.~appreciates useful correspondence with Jens Oluf Andersen. S.K.~acknowledges the hospitality of the Department of Mathematics and Natural Sciences, University of Stavanger, where this work was initiated.


\bibliographystyle{JHEP}
\bibliography{references}


\end{document}